\input harvmac
\let\includefigures=\iftrue
\let\useblackboard=\iftrue
\newfam\black

\includefigures
\message{If you do not have epsf.tex (to include figures),}
\message{change the option at the top of the tex file.}
\input epsf
\def\figin{\epsfcheck\figin}\def\figins{\epsfcheck\figins}
\def\epsfcheck{\ifx\epsfbox\UnDeFiNeD
\message{(NO epsf.tex, FIGURES WILL BE IGNORED)}
\gdef\figin##1{\vskip2in}\gdef\figins##1{\hskip.5in}
instead
\else\message{(FIGURES WILL BE INCLUDED)}%
\gdef\figin##1{##1}\gdef\figins##1{##1}\fi}
\def\DefWarn#1{}
\def\figinsert{\goodbreak\midinsert}
\def\ifig#1#2#3{\DefWarn#1\xdef#1{fig.~\the\figno}
\writedef{#1\leftbracket fig.\noexpand~\the\figno}%
\figinsert\figin{\centerline{#3}}\medskip\centerline{\vbox{
\baselineskip12pt\advance\hsize by -1truein
\noindent\footnotefont{\bf Fig.~\the\figno:} #2}}
\endinsert\global\advance\figno by1}
\else
\def\ifig#1#2#3{\xdef#1{fig.~\the\figno}
\writedef{#1\leftbracket fig.\noexpand~\the\figno}%
\global\advance\figno by1} \fi

\def\id{{1 \kern-.28em {\rm l}}}

\def\K3{{\bf K3}}
\def\journal#1&#2(#3){\unskip, \sl #1\ \bf #2 \rm(19#3) }
\def\andjournal#1&#2(#3){\sl #1~\bf #2 \rm (19#3) }

\def\bar{\overline}

\def\ie{{\it i.e.}}
\def\eg{{\it e.g.}}

\def\tilde{\widetilde}

\def\frac#1#2{{#1\over#2}}

\def\half{\frac12}

\def\inbar{\,\vrule height1.5ex width.4pt depth0pt}
\def\IC{\relax\hbox{$\inbar\kern-.3em{\rm C}$}}
\def\IR{\relax{\rm I\kern-.18em R}}
\def\IP{\relax{\rm I\kern-.18em P}}

%
%

%
\catcode`\@=11
\def\slash#1{\mathord{\mathpalette\c@ncel{#1}}}
\overfullrule=0pt

\def\SS{{\cal S}}

\def\underrel#1\over#2{\mathrel{\mathop{\kern\z@#1}\limits_{#2}}}

\catcode`\@=12


%

\def\det{{\rm det}}

\def\det{{\rm det}}


\lref\PeskinEV{
 M.~E.~Peskin and D.~V.~Schroeder,
 {\it An Introduction to quantum field theory,}
 Reading, USA: Addison-Wesley (1995).

}

\lref\NambuTP{Y.~Nambu and G.~Jona-Lasinio,
``Dynamical Model Of Elementary Particles Based On An Analogy With
Superconductivity. I,''Phys.\ Rev.\  {\bf 122}, 345 (1961).
}

\lref\SakaiCN{
T.~Sakai and S.~Sugimoto,
``Low energy hadron physics in holographic QCD,''
Prog.\ Theor.\ Phys.\  {\bf 113}, 843 (2005)
[arXiv:hep-th/0412141].
}

\lref\GreenDD{
  M.~B.~Green, J.~A.~Harvey and G.~W.~Moore,
  Class.\ Quant.\ Grav.\  {\bf 14}, 47 (1997)
  [arXiv:hep-th/9605033].
}

\lref\KruczenskiBE{
  M.~Kruczenski, D.~Mateos, R.~C.~Myers and D.~J.~Winters,
  JHEP {\bf 0307}, 049 (2003)
  [arXiv:hep-th/0304032].
}

\lref\KruczenskiUQ{
  M.~Kruczenski, D.~Mateos, R.~C.~Myers and D.~J.~Winters,
  JHEP {\bf 0405}, 041 (2004)
  [arXiv:hep-th/0311270].
}

\lref\ItzhakiDD{
N.~Itzhaki, J.~M.~Maldacena, J.~Sonnenschein and S.~Yankielowicz,
``Supergravity and the large N limit of theories with sixteen
supercharges,''
Phys.\ Rev.\ D {\bf 58}, 046004 (1998)
[arXiv:hep-th/9802042].
}

\lref\AntonyanVW{
  E.~Antonyan, J.~A.~Harvey, S.~Jensen and D.~Kutasov,
  ``NJL and QCD from string theory,''
  arXiv:hep-th/0604017.
  }

\lref\GrossJV{
D.~J.~Gross and A.~Neveu,
``Dynamical Symmetry Breaking In Asymptotically Free Field Theories,''
Phys.\ Rev.\ D {\bf 10}, 3235 (1974).
}

\lref\SakaiYT{
T.~Sakai and S.~Sugimoto,
``More on a holographic dual of QCD,''
Prog.\ Theor.\ Phys.\  {\bf 114}, 1083 (2006)
[arXiv:hep-th/0507073].
}

\lref\Witten{
  E.~Witten,
  ``Chiral Symmetry, The 1/N Expansion, And The SU(N) Thirring
Model,''
  Nucl.\ Phys.\ B {\bf 145}, 110 (1978).
}

\lref\Affleck{
  I.~Affleck,
  ``On The Realization Of Chiral Symmetry In (1+1)-Dimensions,''
  Nucl.\ Phys.\ B {\bf 265}, 448 (1986).
}

\lref\ItzhakiTU{
 N.~Itzhaki, D.~Kutasov and N.~Seiberg,
 ``I-brane dynamics,''
 JHEP {\bf 0601}, 119 (2006)
 [arXiv:hep-th/0508025].
}

\lref\pw{ A.~M.~Polyakov and P.~B.~Wiegmann, ``Theory Of
Nonabelian Goldstone Bosons In Two Dimensions,'' Phys.\ Lett.\ B
{\bf 131}, 121 (1983).
}

\lref\ghm{ M.~B.~Green, J.~A.~Harvey and G.~W.~Moore, ``I-brane
inflow and anomalous couplings on D-branes,'' Class.\ Quant.\
Grav.\  {\bf 14}, 47 (1997) [arXiv:hep-th/9605033].
}

\lref\DashenXZ{
R.~F.~Dashen, S.~K.~Ma and R.~Rajaraman,
``Finite Temperature Behavior Of A Relativistic Field Theory With
Dynamical
Symmetry Breaking,''
Phys.\ Rev.\ D {\bf 11}, 1499 (1975).
}

\lref\AharonyDA{
O.~Aharony, J.~Sonnenschein and S.~Yankielowicz,
 ``A holographic model of deconfinement and chiral symmetry
restoration,''
arXiv:hep-th/0604161.
}

\lref\ParnachevDN{
A.~Parnachev and D.~A.~Sahakyan,
``Chiral phase transition from string theory,''
arXiv:hep-th/0604173.
}

\lref\LukyanovNJ{
S.~L.~Lukyanov, E.~S.~Vitchev and A.~B.~Zamolodchikov,
``Integrable model of boundary interaction: The paperclip,''
Nucl.\ Phys.\ B {\bf 683}, 423 (2004)
[arXiv:hep-th/0312168].
}

\lref\LukyanovBF{
S.~L.~Lukyanov and A.~B.~Zamolodchikov,
``Dual form of the paperclip model,''
Nucl.\ Phys.\ B {\bf 744}, 295 (2006)
[arXiv:hep-th/0510145].
}

\lref\GaoUP{
Y.~H.~Gao, W.~S.~Xu and D.~F.~Zeng,
``NGN, QCD(2) and chiral phase transition from string theory,''
arXiv:hep-th/0605138.
}

\lref\KutasovDJ{
D.~Kutasov,
``D-brane dynamics near NS5-branes,''
arXiv:hep-th/0405058.
}

\lref\KutasovRR{
D.~Kutasov,
``Accelerating branes and the string / black hole transition,''
arXiv:hep-th/0509170.
}

\lref\polbook{
  J.~Polchinski,
  ``String theory. Vol. 1: An introduction to the bosonic string,''
  Cambridge, UK: Univ. Pr. (1998).
}

\lref\JacobsYS{
L.~Jacobs,
``Critical Behavior In A Class Of O(N) Invariant Field Theories In
Two-Dimensions,''
Phys.\ Rev.\ D {\bf 10}, 3956 (1974).
}

\lref\HarringtonTF{
B.~J.~Harrington and A.~Yildiz,
``Restoration Of Dynamically Broken Symmetries At Finite
Temperature,''
Phys.\ Rev.\ D {\bf 11}, 779 (1975).
}

\lref\PeetWN{
  A.~W.~Peet and J.~Polchinski,
  ``UV/IR relations in AdS dynamics,''
  Phys.\ Rev.\ D {\bf 59}, 065011 (1999)
  [arXiv:hep-th/9809022].
}
\lref\MyersQR{
  R.~C.~Myers and R.~M.~Thomson,
  ``Holographic mesons in various dimensions,''
  arXiv:hep-th/0605017.
}

\lref\WittenQJ{
  E.~Witten,
   ``Anti-de Sitter space and holography,''
  Adv.\ Theor.\ Math.\ Phys.\  {\bf 2}, 253 (1998)
  [arXiv:hep-th/9802150].
}

\lref\CheungAZ{
  Y.~K.~Cheung and Z.~Yin,
  ``Anomalies, branes, and currents,''
  Nucl.\ Phys.\ B {\bf 517}, 69 (1998)
  [arXiv:hep-th/9710206].
}

\lref\CallanSA{
  C.~G.~.~Callan and J.~A.~Harvey,
  ``Anomalies And Fermion Zero Modes On Strings And Domain Walls,''
  Nucl.\ Phys.\ B {\bf 250}, 427 (1985).
}

\lref\MinasianMM{
  R.~Minasian and G.~W.~Moore,
  ``K-theory and Ramond-Ramond charge,''
  JHEP {\bf 9711}, 002 (1997)
  [arXiv:hep-th/9710230].
}

\lref\KutasovXQ{
  D.~Kutasov and A.~Schwimmer,
  ``Universality in two-dimensional gauge theory,''
  Nucl.\ Phys.\ B {\bf 442}, 447 (1995)
  [arXiv:hep-th/9501024].
}

\lref\tHooftHX{
 G.~'t Hooft,
 ``A Two-Dimensional Model For Mesons,''
 Nucl.\ Phys.\ B {\bf 75}, 461 (1974).
}

\lref\AndreiI{
  N.~Andrei and J.~H.~Lowenstein,
   ``Derivation Of The Chiral Gross-Neveu Spectrum For Arbitrary SU(N)
  Symmetry,''
  Phys.\ Lett.\ B {\bf 90}, 106 (1980).
}

\lref\AndreiII{
  N.~Andrei and J.~H.~Lowenstein,
  ``Diagonalization Of The Chiral Invariant Gross-Neveu Hamiltonian,''
  Phys.\ Rev.\ Lett.\  {\bf 43}, 1698 (1979).
}

\lref\Berg{
  B.~Berg and P.~Weisz,
  ``Exact S Matrix Of The Chiral Invariant SU(N) Thirring Model,''
  Nucl.\ Phys.\ B {\bf 146}, 205 (1978).
}

\lref\Koberle{
  R.~Koberle, V.~Kurak and J.~A.~Swieca,
  ``Scattering Theory And 1/N Expansion In The Chiral Gross-Neveu Model,''
  Phys.\ Rev.\ D {\bf 20}, 897 (1979)
  [Erratum-ibid.\ D {\bf 20}, 2638 (1979)].
}
\lref\toap{E.~Antonyan, J.~A.~Harvey and D.~Kutasov, to appear.}

\Title{\vbox{\baselineskip12pt\hbox{EFI-06-11}
\hbox{}}} {\vbox{\centerline{The Gross-Neveu Model from String Theory}
}}
\bigskip

\centerline{\it E. Antonyan, J.~A. Harvey and D. Kutasov}
\bigskip
\centerline{EFI and Department of Physics, University of
Chicago}\centerline{5640 S. Ellis Av. Chicago, IL 60637}

\smallskip

\vglue .3cm

\bigskip

\bigskip
\noindent We study an intersecting D-brane model which  at low
energies describes $(1+1)$-dimensional chiral fermions localized at
defects on a stack of $N_c$ $D4$-branes. Fermions at different
defects interact via exchange of massless $(4+1)$-dimensional fields.
At weak coupling this interaction gives rise to the Gross-Neveu
(GN) model and can be studied using field theoretic techniques. At
strong coupling one can describe the system in terms of probe branes
propagating in a curved background in string theory. The chiral
symmetry is dynamically broken at zero temperature and is restored
above a critical temperature $T_c$ which depends on the coupling.
The phase transition at $T_c$ is first order at strong coupling and
second order at weak coupling.

\bigskip

\Date{August 2006}

\newsec{Introduction}

In this paper we continue the investigation of dynamical symmetry
breaking in intersecting D-brane models. The construction of these
models starts with $N_c$ BPS $Dp$-branes, which we will refer to as
color branes. The low energy theory on these branes is $(p+1)$-dimensional 
$U(N_c)$ super Yang-Mills (SYM) with sixteen
supercharges. For $p>3$, the Yang-Mills coupling $g_{p+1}$ scales
like a positive power of length, and the SYM degrees of freedom
become free in the infrared.

To get non-trivial infrared dynamics we add additional D-branes,
which we will refer to as flavor branes. The flavor branes intersect
the color ones on subspaces of the $(p+1)$-dimensional worldvolume of
the latter. Open strings stretched between the color and flavor
branes give rise to light fermions (and sometimes bosons as well)
that are localized at the intersections. At low energies, fermions
at different intersections interact via exchange of $(p+1)$-dimensional 
massless fields living on the color branes. In some
cases, these interactions lead to non-trivial infrared effects.

When the $(p+1)$-dimensional $U(N_c)$ SYM theory is weakly coupled in
the infrared, the effective interaction between fermions living at
different defects  becomes weaker as the distance
between the defects, $L$, increases. The leading long distance
dynamics can be studied in an approximation where one includes the
effective four-Fermi interaction that arises from exchange of a
single $(p+1)$-dimensional massless field between two fermions, but
neglects all the higher order exchange processes. Such processes are
suppressed since the gauge coupling $g_{p+1}$ goes to zero at long
distances.

A nice thing about this approximation is that it leads to a theory that
is tractable at
large $N_c$. Indeed, since the only dynamical degrees of freedom are
the fermions,
which transform in the vector representation of $U(N_c)$, the
low-energy dynamics
is described by a {\it vector model}, and can typically be solved using
familiar large
$N_c$ field theoretic techniques.

As $L$ decreases, the interactions between the fermions and those
between the fermions
and the color degrees of freedom become stronger. For small $L$ the
dynamics can in some
cases be analyzed using holography. In this limit the $N_c$ color
branes can be replaced by
their near-horizon geometry, and the interactions among the fermions
are encoded in the
dynamics of the flavor branes in the resulting curved spacetime.

An example of an intersecting brane system of the kind described
above is the $D4-D8-\bar{D8}$ system, which was studied from the
present perspective in \AntonyanVW, following some closely related
earlier work \refs{\SakaiCN,\SakaiYT}. In that model the field
theoretic analysis of the weak coupling regime is complicated by the
fact that the effective four-Fermi interaction due to single gluon
exchange is long range -- it is given by a non-local generalization
of the Nambu-Jona-Lasinio model \NambuTP. This leads to some
interesting subtleties in the above considerations to which we hope
to return elsewhere.

The main purpose of this paper is to analyze an intersecting brane
system which does not
suffer from such subtleties. Our discussion is organized as follows.

In section 2 we describe a brane configuration consisting of $N_c$
color $D4$-branes and $N_f$ flavor $D6$ and $\bar{D6}$-branes in
type IIA string theory, intersecting on a $(1+1)$-dimensional
spacetime. The $D4-D6$ and $D4-\bar{D6}$ intersections are separated
by a distance $L$ in the $\IR^3$ along the fourbranes and transverse
to the sixbranes. At the two intersections one finds chiral left and
right-moving fermions, respectively. The low-energy dynamics of
these fermions is governed by exchange of massless fields that live
on the $D4$-branes, which are described by a five-dimensional gauge
theory with `t Hooft coupling $\lambda$ (which has units of length).
The strength of the interaction among the fermions is determined by
the value of the dimensionless parameter $\lambda/L$.

In section 3 we study the system at weak coupling, where one can (to
leading order in $\lambda/L$) restrict to the single gluon exchange
approximation.  The resulting four-Fermi interaction between the
left and right-moving fermions is described by the Gross-Neveu model
\GrossJV, with a specific UV cutoff. This model is known to break
chiral symmetry, and to generate a mass scale. We rederive the chiral
symmetry breaking from our perspective, and show that the
dynamically generated mass of the fermions is much smaller than
$1/L$. This allows us to take a decoupling limit in which the
intersecting brane model reduces precisely to the GN model.

In section 4 we discuss the system at strong coupling, \ie\ for
$\lambda\gg L$. In this regime the fermions and massless fields on
the $D4$-branes are strongly interacting, but there is still a
weakly coupled description of the physics. It involves replacing the
color $D4$-branes by their near-horizon geometry, and studying the
flavor $D6$-branes as probes in this geometry. As in \AntonyanVW,
there are two possible shapes that the flavor branes can take. One
is the original separated parallel $D6$ and $\bar{D6}$-branes. The
other is a single connected brane which looks like the original $D6$
and $\bar{D6}$-branes connected by a wormhole whose width increases
with $\lambda/L$. We find that (as in \AntonyanVW) the curved,
connected brane has lower energy than the pair of separated straight
ones. This implies that the chiral symmetry acting on the left and
right-moving fermions is dynamically broken in this regime as well.
The energy scale associated with the breaking increases with the
coupling $\lambda/L$. We also comment on the spectrum of the model
in the strong coupling regime and its relation to that of  the
Gross-Neveu model.

In section 5 we discuss the finite temperature behavior of the
model. In the weak coupling regime the chiral condensate decreases
as the temperature increases and vanishes above a certain critical
temperature $T_c$. Thus, the chiral symmetry is restored in a second
order phase transition at $T=T_c$. At strong coupling, the (free)
energy difference between the connected curved and separated
straight branes decreases as the temperature increases, and beyond a
certain critical value of the temperature the straight branes have
lower free energy. The chiral symmetry restoration transition is
strongly first order in this limit.

We end in section 6 with conclusions and a discussion of some open
problems.

\newsec{$D4-D6-\bar{D6}$ brane configuration and its low-energy
dynamics}

The brane configuration that we will consider is depicted in figure 1.
It contains the following $D$-branes:
\eqn\ddconfig{\eqalign{\qquad & 0 ~~~ 1 ~~~ 2 ~~~ 3 ~~~ 4 ~~~5
~~~ 6 ~~~ 7 ~~~ 8 ~~~ 9 ~~~ \cr
D4: ~~& {\rm x} ~~~ {\rm x} ~~~{\rm x} ~~~ {\rm x} ~~~ {\rm x}
~~~{} ~~~ {} ~~~{} ~~~ {} ~~~ {} ~~~ \cr
D6, ~~ \bar{D6}: ~~& {\rm x} ~~~ {\rm x} ~~~ {} ~~~ {}~~~ {} ~~~
{}~~~~~ {\rm x} ~~~ {\rm x} ~~~ {\rm x} ~~~ {\rm x} ~~~ {\rm x} ~~~
\cr
}}
$N_c$ color  $D4$-branes stretched in $(01234)$ are located at the
origin in $(56789)$.
$N_f$ flavor $D6$ and $\bar{D6}$-branes are stretched in $(0156789)$
and separated
by a distance $L$ in $(234)$. Without loss of generality we can take
the separation
between the flavor branes and anti-branes to be in the $x^4$ direction.
The color and
flavor branes intersect on the $(1+1)$-dimensional spacetime labeled by
$(01)$.
We will be primarily interested in the low-energy dynamics at the
intersection.

\ifig\branecon{The configuration of $N_c$ $D4$-branes and $N_f$ $D6$
and
$\bar{D6}$-branes which reduces at weak coupling to the Gross-Neveu
model.}
{\epsfxsize3.0in\epsfbox{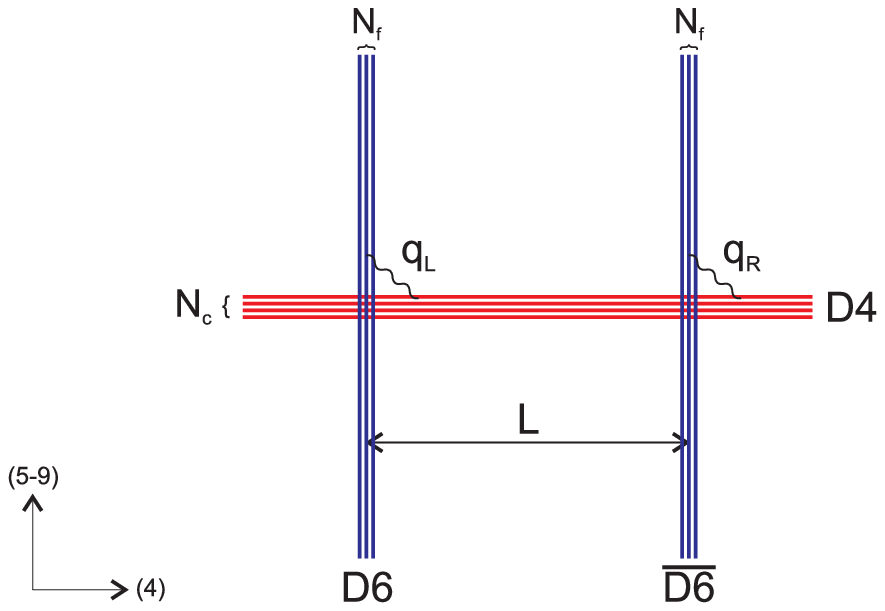}}

The $(9+1)$-dimensional Lorentz symmetry is broken by the branes
\ddconfig\ as follows: \eqn\breaklor{SO(1,9)\to SO(1,1)_{01}\times
SO(2)_{23}\times SO(5)_{56789}~.} Each of the two intersections
preserves a chiral supersymmetry. The $D4-D6$ intersection preserves
eight right-handed supercharges under $SO(1,1)$, while the
$D4-\bar{D6}$ intersection preserves eight left-handed supercharges.
The full system breaks all supersymmetry.

We will study the system of figure 1 in the limit $N_c\to\infty$,
$g_s\to 0$, with $g_sN_c$ and $N_f$ held fixed and $L\gg l_s$. In
this limit the coupling of $6-6$, $\bar 6-\bar 6$ and $6-\bar 6$
strings goes to zero and they become non-dynamical external sources.
The gauge symmetry on the $D6$ and $\bar{D6}$-branes,
\eqn\gaugesym{U(N_f)_L\times U(N_f)_R~,} gives rise to a global
symmetry of the $(1+1)$-dimensional theory at the intersection.

The low-energy degrees of freedom are low lying $4-4$ strings
stretched between color branes, as well as $4-6$ and $4-\bar 6$
strings stretched between color branes and flavor branes and
anti-branes, respectively. The low energy dynamics in the $4-4$
sector is described by dimensional reduction of $(9+1)$-dimensional
$N=1$ SYM with gauge group $U(N_c)$ to $4+1$ dimensions. In the
$4-6$ sector one finds left-moving fermions $q_L$ transforming in
the $(\bar N_f,1)$ of the global symmetry \gaugesym. $4-\bar 6$
strings give right-moving fermions $q_R$ which transform as $(1,\bar
N_f)$. Both $q_L$ and $q_R$ transform in the fundamental $(N_c)$
representation of the $U(N_c)$ gauge symmetry on the $D4$-branes.
Note in particular that the $U(N_f)_L$ symmetry in \gaugesym\  acts
only on the left-moving fermions, while $U(N_f)_R$ acts only on the
right-moving ones. Thus, \gaugesym\ is a symmetry of the Lagrangian,
but we will see later that it is broken to the diagonal $U(N_f)$ by
the dynamics.

The `t Hooft coupling of the $(4+1)$-dimensional gauge theory on the
$D4$-branes, $\lambda$, has units of length. The theory is weakly
coupled for distances much larger than $\lambda$ and strongly
coupled otherwise. In order to define it for all distance scales one
must supply a UV completion. String theory does that by embedding it
in a six-dimensional CFT, the $(2,0)$ theory. The $(4+1)$-dimensional
gauge theory is the low-energy limit of the $(2,0)$ theory
compactified on a circle whose radius is proportional to $g_s l_s$.

The fermions arising from $4-6$ and $4-\bar 6$ strings live on
codimension three defects in the $(4+1)$-dimensional gauge theory.
Their separation, $L$, introduces a second distance scale into the
dynamics. One can think of the defects as probing the dynamics of
the gauge theory (or, more properly, of the $(2,0)$ theory
compactified on $S^1$) at the distance scale $L$.

For $L\gg\lambda$ the theory on the fourbranes is weakly coupled.
Naively, the $(4+1)$-dimensional fields that transform in the adjoint
of $SU(N_c)$ decouple in this limit, just like the corresponding
modes on the sixbranes. This is not quite correct, since exchange of
$(4+1)$-dimensional massless modes leads to a weak attractive
interaction between the left and right-handed fermions for any
finite $\lambda$. As we will see, this interaction leads to chiral
symmetry breaking and has to be kept in the low energy description.

The effective action that describes the dynamics in this limit
contains the fermions $q_L$, $q_R$, and five-dimensional gauge
fields on the $D4$-branes, $A_M$, $M=0,1,2,3,4$, \eqn\lltwo{\SS=\int
d^5x \left[ -{1\over 4g_5^2}F_{MN}^2+ \delta^3(\vec x-\vec x_1)
q_L^\dagger\bar\sigma^\mu(i\partial_\mu+A_\mu)q_L+ \delta^3(\vec
x-\vec x_2) q_R^\dagger\sigma^\mu(i\partial_\mu+A_\mu)q_R \right]
~,} where the notation is as follows. The $(4+1)$-dimensional
worldvolume of the color $D4$-branes is parametrized by
$\{x^M\}=(x^0,x^1,\vec x)$. The flavor branes and anti-branes are
placed at $\vec x=\vec x_1$ and $\vec x_2$, respectively. As
mentioned above, we can take $\vec x_1$, $\vec x_2$ to lie on the
$x^4$ axis, with $|\vec x_1-\vec x_2|=L$. The indices $\mu$ in
\lltwo\ run over the range $\mu=0,1$ or $+,-$. $q_L$, $q_R$ are
complex one component spinors. The corresponding one-dimensional
Dirac matrices are $\bar\sigma^\mu=\delta^\mu_+$,
$\sigma^\mu=\delta^\mu_-$.

We will define the `t Hooft coupling of the $(4+1)$-dimensional gauge
theory on the color branes in terms of $g_5$ \lltwo\ by\foot{Here
and below we often set $\alpha'=1$.}
\eqn\thooft{\lambda=g_sN_c={g_5^2N_c\over(2\pi)^2}~.} As mentioned
above, $\lambda$ is a length scale. In the next section we will use
the Lagrangian \lltwo\ to analyze the interaction between the
fermions $q_L$ and $q_R$ for $\lambda\ll L$. In the opposite limit
of large $\lambda/L$ we cannot use \lltwo, both because we have to
add to it the other massless fields living on the $D4$-branes, and
because in that regime the dynamics of these modes is no longer
described by weakly coupled field theory in $4+1$ dimensions.
Nevertheless, one can still efficiently study the theory in this
limit. This will be discussed in section 4.

\newsec{Weak coupling analysis}

In order to study the dynamics of the fermions $q_L$, $q_R$ at weak
coupling, it is convenient to integrate out the five-dimensional
fields living on the color $D4$-branes. To leading order in
$\lambda/L$ this can be done by taking into account the interaction
among the fermions due to single gluon exchange, using the action
\lltwo. The calculation is similar to that of \AntonyanVW\ and leads
to
\eqn\lreff{\eqalign{
\SS_{\rm eff}= & i\int d^2x \left(q_L^\dagger\bar\sigma^\mu\partial_\mu
q_L
+q_R^\dagger\sigma^\mu\partial_\mu q_R\right) \cr
& ~~~ +{g_5^2 \over 4 \pi^2} \int d^2x d^2y G(x-y,L) \left(
q^\dagger_L(x)\cdot q_R(y) \right)
\left( q^\dagger_R(y)\cdot q_L(x) \right) \cr
}}
As in \AntonyanVW,  $q^\dagger_L(x)\cdot q_R(y)$ is a global color
singlet (hence the dot) and transforms in the $(N_f, \bar N_f)$ of
$U(N_f)_L\times U(N_f)_R$. $G(x,L)$ is the five-dimensional massless
propagator over a distance $x$ in the directions $(01)$ and $L$ in
the directions $(234)$, \eqn\fivedprop{G(x,L)={1\over
(x^2+L^2)^{3\over2}}~.} In addition, there are self interactions
between fermions at a given intersection (the analog of eq. (3.2) in
\AntonyanVW), but these are small for distances much larger than
$\lambda$ and can be neglected. We also Wick rotated to Euclidean
space, which is convenient for analyzing the vacuum structure.

Many properties of the solution of the interacting quantum field
theory corresponding to the action \lreff\ can be anticipated as
follows. The main complication in \lreff\ is the non-local four
Fermi interaction on the second line. If the function $G(x-y,L)$ was
proportional to $\delta^2(x-y)$, \lreff\ would reduce to the
Gross-Neveu model \GrossJV, an asymptotically free local quantum
field theory which is exactly solvable in the large $N_c$ limit.

The actual Green function, \fivedprop\ can be thought of as a
$\delta$-function smeared over a region of size $L$. Using the fact
that \eqn\intgx{\int d^2xG(x,L)={2\pi\over L}~,} we can, at length
scales large compared to $L$,  replace \lreff\ by the local action
\eqn\lgn{ \SS_{\rm eff}= \int d^2x
\left[iq_L^\dagger\bar\sigma^\mu\partial_\mu q_L
+iq_R^\dagger\sigma^\mu\partial_\mu q_R +{1\over N_c}{2\pi\lambda
\over L}  \left( q^\dagger_L(x)\cdot q_R(x) \right) \left(
q^\dagger_R(x)\cdot q_L(x) \right) \right]}
with a UV cutoff $\Lambda\simeq{1\over L}$. Comparing to \GrossJV\
we see that the Gross-Neveu coupling $\lambda_{gn}$ is given in
terms of our parameters by
\eqn\lamgn{\lambda_{gn}={2 \pi\lambda\over L}~.}
In \GrossJV\ it was found that at large $N_c$ the model \lgn\
dynamically generates a fermion mass which goes like
\eqn\massferm{M_F\simeq \mu e^{-{2 \pi\over\lambda_{gn}}}}
where $\mu$ is an arbitrary renormalization scale and $\lambda_{gn}$
the coupling at that scale. Since our model reduces in a particular
limit (which will be made more precise below) to the Gross-Neveu
model, it is natural to expect that it exhibits similar behavior. We
next verify this by a more detailed analysis following closely that
of \AntonyanVW.

To solve the theory \lreff\ for large $N_c$ it is convenient to
introduce a bilocal field $T(x,y)$ which is a singlet of global
$U(N_c)$ and transforms as $(N_f,\bar N_f)$ under the 
symmetry \gaugesym. The action \lreff\ can be rewritten as
\eqn\tlreff{\eqalign{
\SS_{\rm eff} = i&\int d^2x \left(q_L^\dagger\bar\sigma^\mu\partial_\mu
q_L
+q_R^\dagger\sigma^\mu\partial_\mu q_R\right)\cr
+&\int d^2x d^2y \left[-{N_c\over\lambda}{T(x,y)\bar T(y,x)\over
G(x-y,L)}+
\bar T(y,x)q^\dagger_L(x)\cdot q_R(y)+T(x,y)q^\dagger_R(y)\cdot
q_L(x)\right]~.
\cr}}
Integrating out $T(x,y)$ using its equation of motion,
\eqn\eomtt{T(x,y)={\lambda\over N_c}G(x-y,L)q_L^\dagger(x)\cdot
q_R(y)~,}
one recovers the original action \lreff. To solve the large $N_c$
theory one instead integrates out the fermions and obtains an
effective action for $T$, which becomes classical in the large $N_c$
limit. As in \AntonyanVW, using translation invariance of the
vacuum, it is enough to compute this effective action for
$T(x,y)=T(|x-y|)$. Dividing by $N_c$ and the volume of spacetime we
get the effective potential  \eqn\gapa{ V_{\rm eff}=
{1\over\lambda}\int d^2 x {T(x) \bar T(x) \over G(x,L)} - \int {d^2
k \over (2 \pi)^2} \ln\left(1 + {T(k) \bar T(k) \over k^2}\right) }
where $T(k)$ is the Fourier transform of $T(x)$. Varying \gapa\
with respect to $\bar T(k)$ leads to the gap equation
\eqn\gapeq{{1\over\lambda}\int d^2x {T(x)\over G(x,L)}e^{-ik\cdot x}
={T(k)\over k^2+T(k)\bar T(k)}~.}
The solution of \gapeq\ gives the expectation value of the chiral
condensate $\langle q_L^\dagger(x)\cdot q_R(y)\rangle$ via \eomtt. 
This equation has a trivial solution $T=0$, but we will next see that 
there is a non-trivial solution as well. As shown in \AntonyanVW, any 
well behaved non-trivial solution of \gapeq\ has a negative value of 
$V_{\rm eff}$; hence the vacuum of the theory breaks chiral symmetry.

To solve the gap equation for small $\lambda/L$ it is
convenient to define the field $f(x,y)$ by
\eqn\mnonlineans{T(x,y)={L\over2\pi}G(x-y,L)f(x,y)~.}
Using \eomtt\ we see that the expectation value of $f(x,y)$ is
proportional to the condensate $\langle q^\dagger_L(x)\cdot
q_R(y)\rangle$, 
\eqn\fqq{f(x-y)\equiv\langle
f(x,y)\rangle={\lambda_{gn}\over N_c}\langle q_L^\dagger(x)\cdot
q_R(y)\rangle~.} In particular, the condensate at coincident points,
$\langle q_L^\dagger(x)\cdot q_R(x)\rangle$ is proportional to
$f(0)$, which we will denote by $m_f$. We will determine $m_f$ as
part of the solution of the gap equation. We will assume (and
justify later) that $f(x)$ is approximately constant on the scale
$L$. Since the function $G(x,L)$ \fivedprop\ goes rapidly to zero
for $x>L$ (and can be thought of as a smeared $\delta$-function)
this implies that
\eqn\mmomspace{T(k)= {L\over2\pi}\int d^2x G(x,L)f(x)e^{-ik\cdot
x}\simeq m_f e^{-kL} ~.}
Substituting \mnonlineans\ and \mmomspace\ into the gap equation
\gapeq\ gives the Fourier
transform of $f(x)$:
\eqn\mnewgap{{\tilde f}(k) = \lambda_{gn} {m_f e^{-|k| L} \over k^2 +
m_f^2 e^{-2 |k| L}}~.}
We can now determine the mass parameter $m_f$ by evaluating $f(0)$  in
terms of its
Fourier transform:
\eqn\lmmfft{m_f  = f(0)= \lambda_{gn} \int {d^2 k \over (2 \pi)^2}
{m_f e^{-|k| L} \over k^2 + m_f^2 e^{-2 |k| L}}~.}
Dividing by $m_f$ and estimating the momentum integral for $m_fL\ll 1$
we find
\eqn\mmans{1 \simeq {\lambda_{gn} \over 2 \pi} \log(\Lambda/m_f) }
where $\Lambda\sim 1/L$ is the UV cutoff of the effective GN model
\lgn. Note
that the dynamically generated mass scale
\eqn\mmferm{m_f \simeq \Lambda e^{-2 \pi/\lambda_{gn}}}
is much smaller than the cutoff $\Lambda$ for small GN coupling.

To complete the discussion, we need to show that the resulting $f(x)$
is slowly varying on the scale $L$ (to justify \mmomspace). Fourier
transforming \mnewgap\ we find the following behavior. For $x\ll L$,
\eqn\xlessl{f(x)\simeq f(0)=m_f~.}
For $L\ll x\ll 1/m_f$,
\eqn\xmorel{f(x)\simeq m_f{\ln(xm_f)\over \ln(Lm_f)}~,}
while for $x\gg1/m_f$, $f$ goes exponentially to zero.

Although $f(x)$ for $x>L$ is not constant, it is very slowly varying at
weak coupling.
Indeed,
\eqn\vrafx{{f(0)-f(x)|_{x \gg L}\over f(0)}\simeq
1-{\ln(xm_f)\over\ln(Lm_f)}\simeq
{\lambda_{gn}\over2\pi}\ln{x\over L}~.}
The logarithmic variation \xmorel, \vrafx\ becomes important only on
scales for
which ${\lambda_{gn}\over2\pi}\ln{x\over L}\simeq 1$, \ie\ $x\simeq
1/m_f$. We
conclude that $f(x)$ \mnonlineans\ varies on the scale $1/m_f\gg L$.

The same conclusion can be reached by examining the momentum space
expression \mnewgap.
The Fourier transform ${\tilde f}(k)$ contains two momentum scales,
$m_f$ and $1/L$. However,
most of its support is on the scale $m_f$. At $k=1/L$ it has been
reduced by a factor
$e^{-2L/\lambda}$ from its value at the origin. This means that the
scale of variation
of $f(x)$ is $1/m_f$.

The solution for $f$ gives the following position dependent
condensate:
\eqn\poscond{\langle q_L^\dagger(x)\cdot
q_R(0)\rangle=N_cm_f\int_{|k|<\Lambda}
{d^2k\over(2\pi)^2}{e^{ik\cdot x}\over k^2+m_f^2} ~.}
This is nothing but the massive propagator for a fermion of mass $m_f$.
Thus, we
conclude that the intersecting D-brane model generates dynamically a
mass $m_f$
\mmferm\ for the fermions living at the intersections of the color and
flavor branes.
This mass is non-perturbative in the Gross-Neveu coupling \lamgn.
Comparing
\massferm\ to \mmferm\  we see that is exhibits the same dependence on
the coupling
as in the field theoretic analysis of \GrossJV.

One can take a decoupling limit  in which the physics of the brane
configuration reduces
precisely to the Gross-Neveu model. To do that one takes $\lambda/L\to
0$ and focuses
on energy scales of order $m_f$.  This sends the ratio of the mass of
the fermions $m_f$
\mmferm\ and the UV cutoff scale $1/L$ to zero, and leads to a model
that is precisely
equivalent to that of \GrossJV. The model with finite $\lambda/L$  can
be thought of as a
version of Gross-Neveu with a finite UV cutoff.

In the above discussion we took the renormalization scale of the theory
to be the UV cutoff $\Lambda \sim 1/L$.  $\lambda_{gn}$ \lamgn\ is the 
coupling at that scale. One can choose some other renormalization scale 
$\mu < \Lambda$ and define the coupling $\lambda_{gn}$ at that scale. 
The renormalization group guarantees that Green functions expressed in 
terms of the physical fermion mass are insensitive to such changes.

The Gross-Neveu model \lgn, which describes the dynamics of $q_L$
and $q_R$ at weak coupling, was studied extensively in the past, and
much is known about it. We next review some of its pertinent
features. We first address the question of its extreme low-energy
limit. This can be analyzed as follows (see \eg\
\refs{\Witten,\Affleck}). The CFT of free massless fermions $q_L$,
$q_R$ can be bosonized in terms of a WZW model for the group
\eqn\wzwgroup{SU(N_c)_{N_f}\times SU(N_f)_{N_c}\times U(1)~.}
The subscripts in \wzwgroup\ denote the levels of the corresponding
current algebras. One can write the left-moving currents
corresponding to \wzwgroup\ in terms of the fermions $q_L$ as
follows: \eqn\currents{\eqalign{ U(1):&\qquad
J=(q_L^*)^\alpha_i(q_L)^i_\alpha~;\cr SU(N_c)_{N_f}:&\qquad
J^\alpha_\beta=(q_L^*)^\alpha_i(q_L)^i_\beta-{1\over
N_c}\delta^\alpha_\beta J~;\cr SU(N_f)_{N_c}:&\qquad
J^i_j=(q_L^*)^\alpha_j(q_L)^i_\alpha-{1\over N_f}\delta^i_j J~.\cr
}} Here $\alpha,\beta=1,\cdots, N_c$ are color indices and
$i,j=1,\cdots, N_f$ are flavor ones. A similar decomposition holds
for the right-moving fermions. The currents on each line of
\currents\ satisfy standard affine Lie algebra OPE's, and commute
with those on the other two lines.  The $U(1)$ current on the first
line describes a massless free field. The rest of the low-energy
theory \wzwgroup\ is an interacting WZW model (for generic $N_f$,
$N_c$).

The four-Fermi interaction \lgn\ can be written in terms of the
currents \currents, as a linear combination of an abelian Thirring
interaction for the $U(1)$ factor, and a non-abelian Thirring
interaction corresponding to $SU(N_c)$. The abelian Thirring
interaction \eqn\uone{\CL_1=\lambda_1 J\bar J} corresponds to
changing the radius of the decoupled scalar field associated with
the $U(1)$ factor in \wzwgroup. Thus, its effects are easy to
analyze. The non-abelian Thirring interaction associated with
$SU(N_c)$, \eqn\sunc{\CL_2=\lambda_2J^\alpha_\beta \bar
J_\alpha^\beta~,} leads to non-trivial long distance dynamics. The
coupling $\lambda_2$ is asymptotically free. It grows in the
infrared, where the $SU(N_c)_{N_f}$ degrees of freedom become
massive and decouple. The extreme infrared theory is a coset model.
It is obtained from the UV theory, which contains the current
algebra \currents, by gauging the $SU(N_c)_{N_f}$ subalgebra. This
eliminates the corresponding factor in \wzwgroup\ and leads to a WZW
model for \eqn\extremeir{SU(N_f)_{N_c}\times U(1)~.} One can think
of this model as describing the infrared fluctuations of the order
parameters associated with chiral symmetry breaking.

As discussed in  \refs{\Witten,\Affleck} these fluctuations
effectively restore the chiral symmetry at finite $N_c$. For
example, the two point function \eqn\twopt{\langle
q^\dagger_L(x)\cdot q_R(x) q^\dagger_R(y)\cdot q_L(y)\rangle} does
not in fact approach a constant at large separation, as would be
expected from \poscond, but rather decays like $|x-y|^{-{1\over
N_c}}$.  Thus, at infinite $N_c$ the symmetry is broken, but at
finite $N_c$ it is restored at sufficiently long distances. As $N_c$
increases, the range of distances for which the symmetry appears to
be broken increases exponentially with $N_c$, and the large $N_c$
analysis becomes a better and better approximation to the dynamics.

It will be useful below to determine the location of the massless
modes \extremeir\  in the extra dimensions. This can be done
following \ItzhakiTU\ by integrating out the fermions $q_L$, $q_R$
and studying the normalizable modes of the resulting theory of gauge
fields. To determine the fate of the $SU(N_f)$ factor\foot{The
$U(1)$ factor in \extremeir\ can also be treated as in \ItzhakiTU.}
in the WZW model \extremeir\ it is enough to consider a single
intersection, say that of the $N_c$ $D4$-branes and $N_f$
$D6$-branes. Since we are interested in the low-energy dynamics, we
can restrict to s-waves on the four-spheres in the $(56789)$
directions and write an effective three-dimensional Lagrangian for
the three-dimensional $SU(N_f)$ gauge field with components\foot{$u$
is the radial direction in $(56789)$.} $(A_+,A_-,A_u)$. That
Lagrangian contains three types of terms,
\eqn\fulll{\CL=\CL_{\rm kin}+\CL_{\rm ferm}+\CL_{\rm inflow}~.}
The first is the standard kinetic term, which has the form
\eqn\lkin{\CL_{\rm kin}={1\over g_7^2} u^4{\rm
Tr}\left[\half(F_{+-})^2-F_{u+}F_{u-}\right]~.} The second  comes
from integrating out the fermions at the intersection. It is
localized at $u=0$ and has the form \eqn\lferm{\CL_{\rm
ferm}=\delta(u)N_c\Gamma(A(0))~.} Here $\Gamma(A)$ is the chiral
Polyakov-Wiegmann action \pw\ corresponding to $SU(N_f)$, and $A(0)$
stands for the two-dimensional gauge field at the intersection,
$A_\pm(x^+,x^-,0)$.

The third term in \fulll\ is a Chern-Simons term, which arises as
follows. On the worldvolume of the $D6$-branes there is a coupling
of the $SU(N_f)$ gauge field $A$ to the RR four-form field
strength\foot{We normalize the RR forms and Chern-Simons couplings
as in \SakaiCN.} $H_4$: \eqn\ccss{{1\over8\pi^2}\int H_4\wedge
\omega_3(A)} where $\omega_3(A)$ is the Chern-Simons form for
$SU(N_f)$, \eqn\cherns{\omega_3(A)={\rm Tr}\left(A\wedge
F+{2\over3}A\wedge A\wedge A\right)~.} One way of deducing the
presence of the term \ccss\ and its coefficient is to require the
cancellation of the $SU(N_f)$ anomaly of the fermions living on the
$D4-D6$ intersection by anomaly inflow \ghm.

The $N_c$ $D4$-branes \ddconfig\ are localized in the directions
$(56789)$ and are magnetically charged under $H_4$. Therefore, there
are $N_c$ units of $H_4$ flux going through the four-sphere at fixed
$u$,
\eqn\hfoursrc{\int_{S^4} H_4 = 2 \pi N_c~. }
Integrating \ccss\ over the sphere leads to a Chern-Simons term in
the Lagrangian \fulll: \eqn\linflow{\CL_{\rm inflow}={N_c\over 4
\pi}\omega_3(A)~.} The resulting Lagrangian \fulll\ is very similar
to one that was recently studied in \ItzhakiTU. An analysis similar
to that of \ItzhakiTU\ shows that the $SU(N_f)$ degrees of freedom,
which classically live at the intersection $u=0$, are displaced to
$u\simeq \lambda^{1\over3}$. This displacement is small for weak
coupling, but it has interesting implications at strong coupling, as
in \ItzhakiTU.

Note that while in the above discussion we analyzed the position of
the chiral $SU(N_f)$ modes associated with a given intersection, the
answer is the same for the system we are interested in, which has 
two intersections. As explained above, in that system there is  a 
Thirring interaction for the $SU(N_c)$ and $U(1)$ factors in \wzwgroup. 
Since the $SU(N_f)$ currents commute with these perturbations, they 
are not influenced by them. This decoupling between the massless sector 
and a non-trivial massive sector of a two-dimensional field theory is  a rather 
general phenomenon; see \eg\  \KutasovXQ\ for a discussion.

So far we focused on the properties of the massless modes associated
with the brane configuration of section 2. The massive spectrum of
the non-abelian Thirring model \sunc\ is also known, since the model
is integrable \refs{\Berg\AndreiI\AndreiII-\Koberle}. The masses of single particle states are given by
\eqn\masses{M_n=m_f{\sin{\pi n\over N_c}\over\sin{\pi\over N_c}}~.}
The integer $n$ runs over the range $n=1,2,\cdots, N_c-1$. The $n=1$
states are the original fermions. States with higher $n$ can be
thought of as bound states of $n$ fermions.

\newsec{Strong coupling analysis}

Bringing the $D6$-branes closer or increasing $g_sN_c$ increases
$\lambda/L$ and takes
us to the strong coupling regime in which the  approximations used in
the previous section
break down. For $L\ll \lambda$ we can employ a dual description of the
model by studying
the DBI action for $D6$-branes in the near-horizon geometry of the
$N_c$ $D4$-branes. The
discussion follows closely that of \refs{\AntonyanVW,\SakaiCN}.

\subsec{Chiral Symmetry Breaking}

The near-horizon metric and dilaton of the $D4$-branes are given by
\eqn\dfour{\eqalign{&ds^2=\left( U \over R\right)^{3\over2}
\left[ \eta_{\mu \nu} dx^\mu dx^\nu-(d x^4)^2\right]-
\left(U\over R\right)^{-{3\over2}}
\left(dU^2+U^2d\Omega_4^2\right)~,\cr
&e^\Phi=g_s\left(U\over R\right)^{3\over4}~,}}
where we set $\alpha'=1$, $\Omega_4$ labels the angular directions in
$(56789)$,
and the parameter $R$ is given by
\eqn\rdefn{R^3 =  \pi g_s N_c =
{g_5^2 \over 4 \pi} N_c =  \pi \lambda~.}
Recall that the $(4+1)$-dimensional `t Hooft coupling $\lambda$ has
units of length. There is also a non-zero RR four-form field strength 
$H_4$ with $N_c$ units of flux around the $S^4$ in \dfour.

The $D6$-brane\foot{We write all the formulae for $N_f=1$, but it is
easy to generalize them to larger $N_f$.} is stretched in the
directions (01), wraps the four-sphere labeled by $\Omega_4$, and
forms a curve $U=U(x^4)$ in the $(U,x^4)$ plane. The induced metric
on it is
\eqn\indmet{d \tilde s^2 = \left( U \over R\right)^{3\over2} \left[
(dx^0)^2 - (dx^1)^2 \right]
- \left( U \over R\right)^{3\over2} \left[ 1+ \left( U \over
R\right)^{-3} U'^2 \right] (dx^4)^2
- \left( U \over R\right)^{-{3\over2}} U^2 d \Omega_4^2}
where $U'=dU/dx^4$. The curve $U(x^4)$ is determined by solving the
equations of motion that follow from the DBI action for the sixbrane
\eqn\dbi{S_{D6} =-\tau_6 V_{1+1}V_4 R^{3\over2}\int dx^4 U^{5\over2}
\sqrt{1+ \left(R\over U\right)^3 U'^2}~.}
$V_{1+1}$ is the volume of $(1+1)$-dimensional Minkowski spacetime, $V_4$
is the volume
of a unit $S^4$, and $\tau_6 = T_6/g_s$ is the tension of a $D6$-brane
for constant dilaton
$e^{\Phi}=g_s$.

Since the Lagrangian does not depend explicitly on $x^4$, there is a
first integral given by
\eqn\conserv{{U^{5\over2} \over \sqrt{1 + \left( R \over U \right)^3
U'^2}} = U_0^{5\over2}}
where $U_0$ is the value of $U$ where $U'=0$. We are looking for
solutions which asymptotically approach a $D6$ and $\bar{D6}$-brane
a distance $L$ apart, \ie\ $U(x^4\to\pm{L\over2}) \to\infty$. $U_0$
is the smallest value of $U$ along the brane.

There are two types of such solutions. The first is a straight brane
and anti-brane stretched in $U$ and located at $x^4=\pm{L\over2}$, 
corresponding\foot{Since the curvature diverges as $U\to 0$, the small
$U$ behavior of this solution is not reliably described by supergravity. 
This is not going to matter for our considerations that will mainly involve
the large $U$ region.}  to infinite $U'$ and $U_0=0$ in \conserv. This 
solution preserves the chiral $SU(N_f)_L\times SU(N_f)_R$ symmetry.

The second is a symmetric U-shaped curve in the $(U,x^4)$ plane.
Integrating \conserv,
it is given by
\eqn\dbisln{x^4(U) = \pm\int_{U_0}^U {du \over \left( {u \over R}
\right)^{3\over2} \left( {u^5 \over U_0^5} -1 \right)^{1\over2}} =
\pm{1 \over 5} {R^{3\over2} \over U_0^{1\over2}}
\left[B\left({3\over5},{1\over2}\right) - B\left({U_0^5\over U^5};
{3\over5},{1\over2}\right) \right]}
where $B(z;a,b)$ is the incomplete Beta function, with $B(0;a,b)=0$
and $B(1;a,b)=B(a,b)$ the usual Euler Beta function. The brane
separation is \eqn\L{L=2x^4(\infty)={2 \over 5} {R^{3\over2} \over
U_0^{1\over2}}B\left({3\over5},{1\over2}\right)~.}
This solution can be thought of as a $D6$-brane and a
$\bar{D6}$-brane connected by a wormhole, whose radius at its
narrowest point is $U_0$. The fact that what looks asymptotically
like $D6$ and $\bar{D6}$-branes  is in fact part of a single
connected curved brane implies that the chiral $U(N_f)_L\times
U(N_f)_R$ symmetry is broken down to the diagonal $U(N_f)$.

To find the ground state of the system one has to compare the energies
of the straight and curved solutions for fixed  $L$. The energy
difference
$\Delta E \equiv E_{\rm straight}-E_{\rm curved}$ is proportional to
\eqn\endiff{
\Delta {E}  \sim
  \int_0^{U_0} du\left( u - 0 \right)
  + \int_{U_0}^\infty du u\left[ 1 -
\left(1-\frac{U_0^5}{u^5}\right)^{-{1\over2}} \right]
= -\frac{1}{5}U_0^{2} B\left(-{2\over5},{1\over2}\right) \approx
0.139\, U_0^2 \;, }
so the curved configuration is preferred and chiral symmetry is broken.
We
see that the system is in the same phase for weak and  strong
coupling.

The regime of validity of the supergravity approximation is the same as
in
\refs{\SakaiCN,\AntonyanVW},  $\lambda\gg L$ or large effective `t
Hooft
coupling, $g_s(U_0)N_c\gg 1$, \ItzhakiDD.

\subsec{Spectrum}

The wormhole connecting the $D6$ and $\bar{D6}$-branes in the vacuum
configuration described in the previous subsection provides a mass
to the fermions $q_L,q_R$.  These correspond at strong coupling to
fundamental strings that stretch from the bottom of the curved
flavor branes down to $U=0$.\foot{The fact that the lowest lying
state of such a string is a fermion can be understood as follows.
The background \dfour\ is close to flat except very close to $U=0$.
The open string in question ends on one side on a brane which is
stretched in the directions (01), $x^4$ and an $S^4$ of radius
$U_0$. The other end of the string is at $U=0$, where the curvature
is large, but  we can replace this region by
boundary conditions corresponding to an open string ending on
$D4$-branes extended in (01234). Overall, the string has six
Dirichlet-Neumann directions (the directions (23) and the $S^4$), so
the lowest lying state is a (massive) fermion.} Using the metric
\dfour\ one finds that their mass is equal to $U_0/2\pi\alpha'=U_0/2\pi$, which
from \L\ is proportional to $\lambda/L^2$. This should be compared
to the weak coupling regime, where we saw that the dynamically
generated mass of the fermions \mmferm\ was given by $m_f \sim (1/L)
e^{-L/\lambda}$. Defining
\eqn\massll{m_f L = A(\lambda/L)~,}
$A(x)$ behaves as
\eqn\aaxx{A(x) \simeq \cases{e^{-1/x}, & \qquad $x \rightarrow 0$ ;
\cr
                                                      x, & \qquad $x
\rightarrow \infty$. \cr } }
For $x\simeq 1$ both our approximations break down, but it is
natural to expect $A(x)$ to be monotonic in $x$, so that the
dynamically generated mass is a monotonic function of the coupling.

In addition to the long strings discussed above the spectrum
includes light mesons which correspond to normalizable fluctuations
of the scalars, gauge field and fermions on the $D6$-brane. To study
them we need to expand the DBI action around the background
configuration \dbisln.  In the $D4-D8-\bar{D8}$ system  it was shown
in \SakaiCN\  that fluctuations of the gauge field give rise to
massless Nambu-Goldstone (NG) bosons and massive vector mesons,
while scalar fluctuations lead to a spectrum of massive scalar
mesons.  We will see that in our model gauge field fluctuations do
not give rise to normalizable NG modes. Also, while the kinetic term
of the gauge field gives rise to  a massless vector,  it acquires a
mass by mixing with other modes via the Chern-Simons interaction
\ccss.

We start with the action for the probe $D6$-brane including the
worldvolume gauge field:
\eqn\dsixgauge{ S_{D6} = - T_6 \int d^7x
e^{-\Phi}\sqrt{-\det(\tilde{g}_{AB}+2\pi
F_{AB})}+ \frac{1}{8\pi^2} \int H_4 \wedge \omega_3(A) \; .}
Here $H_4$ is the RR field sourced by the color $D4$-branes (as in
\ccss) and $\omega_3$ is the Chern-Simons three-form \cherns. In the
four-dimensional model of \SakaiCN\ there is an analogous coupling
to the Chern-Simons five-form, which plays an important role in
anomaly cancellation, but since it is cubic in the gauge field, it
does not affect the mass spectrum. In our case,  the Chern-Simons
term contains terms quadratic in the gauge field and must be
included in the analysis.

We focus on $SO(5)$ singlet gauge field fluctuations that are
independent of the $S^4$
coordinates. Denoting the remaining components of the gauge field by
$A_m$, $m=0,1,U$
and expanding the action to quadratic order we find
\eqn\dsixflucs{S_{D6}^{\rm (quad)}  = -T_6 V_4(2\pi)^2  \int d^2xdU
e^{-\Phi}
\sqrt{-\det{\tilde{g}}}\frac{1}{4}F_{mn} F^{mn} + \frac{N_c}{4\pi}
\int
d^2xdU\epsilon^{mnp}A_m F_{np} ~.}
In this expression $\det \tilde g$ is the determinant of the induced
metric on the $D6$-brane
\indmet, and the  indices $m,n$ are raised and lowered with the $m,n$
components of this
metric. Using the solution \dbisln\ we can rewrite the action
\dsixflucs\ as
\eqn\dsixnext{\eqalign{ S_{D6}^{\rm (quad)} = - T_6 V_4(2\pi)^2  &\int
d^2xdU
\Bigl[\frac{1}{4} G(U) F^{\mu \nu} F_{\mu \nu} + \frac{1}{2} H(U)
{F^\mu}_U F_{\mu
U}\Bigr] \cr
+ \frac{N_c}{4\pi} &\int d^2xdU \epsilon^{\mu \nu} \left( A_U F_{\mu
\nu} + 2
A_\mu F_{\nu U} \right) \; .\cr}}
In \dsixnext\ the indices $\mu,\nu=0,1$ are raised and lowered with the
flat metric
$\eta_{\mu \nu}$ and all $U$ dependence is contained in the functions
\eqn\ghdefs{\eqalign{G(U) & = {R^6 \gamma(U)^{1\over2} \over g_s
U^2} ~,\cr
                                       H(U) & = {R^3 U \over g_s
\gamma(U)^{1\over2}} ~,\cr
                                       \gamma(U) & = {U^5 \over U^5 -
U_0^5} ~.\cr }}
We now separate variables and expand the gauge field in modes in the
$U$ direction as
follows (with normalization conditions to be determined shortly)
\eqn\gaugemodes{\eqalign{
&A_{\mu}(x^{\nu},U)=\sum_n B_{\mu}^{(n)}(x^{\nu})\psi_n(U) \;, \cr
&A_{U}(x^{\nu},U)=\sum_n \pi^{(n)}(x^{\nu})\phi_n(U) \;.\cr
}}
Defining $B_{\mu \nu}^{(n)} = \partial_\mu B_\nu^{(n)} - \partial_\nu
B_\mu^{(n)}$ this leads to the
action
\eqn\expact{\eqalign{S=& -T_6 V_4 (2 \pi)^2 \int d^2x dU  \Bigl[
\sum_{n,m} G(U) \psi_n \psi_m
{1 \over 4} B_{\mu \nu}^{(n)} B^{(m) \mu \nu}  \cr
 & + {1 \over 2} H(U) \sum_{n,m} \left( \partial_\mu \pi^{(n)} \phi_n -
B_\mu^{(n)} \partial_U \psi_n \right)
 \left( \partial^\mu \pi^{(m)} \phi_m - B^{(m) \mu} \partial_U \psi_m
\right)\Bigr] \cr
& + {N_c \over 4\pi} \int d^2x dU \Bigl[\epsilon^{\mu \nu} \sum_{n,m}
\left( \pi^{(n)} \phi_n B_{\mu
\nu}^{(m)} \psi_m + 2 B_\mu^{(n)} \psi_n \left( \partial_\nu \pi^{(m)}
\phi_m - B_\nu^{(m)} \partial_U
\psi_m \right) \right) \Bigr] \; .}}
We see from the first line of \expact\ that the kinetic  term for
$B_\mu^{(n)}$ is canonical provided that  the wavefunctions 
$\psi_m(U)$ satisfy the normalization condition
\eqn\psinorm{T_6 V_4 (2 \pi)^2 \int dU G(U) \psi_n(U) \psi_m(U) =
\delta_{n,m}~.}
Terms on the second line of \expact\ give diagonal and canonically
normalized mass terms for
$B_\mu^{(n)}$ with mass squared $m_n^2$ provided that we also have
\eqn\psinormtwo{T_6 V_4 (2 \pi)^2 \int dU H(U) \partial_U \psi_n(U)
\partial_U \psi_m(U) =
m_n^2 \delta_{n,m} \; .}
Equations \psinorm\ and \psinormtwo\ are compatible if $\psi_n$
satisfies the
eigenvalue equation
\eqn\psieigen{{1 \over G(U)} \partial_U \left[ H(U) \partial_U
\psi_n(U) \right] = - m_n^2 \psi_n(U) \; .}
Finally, in order to have canonically normalized kinetic energy terms
for the $\pi^{(n)}$
we require
\eqn\phinorm{T_6 V_4 (2 \pi)^2 \int dU H(U) \phi_n(U) \phi_m(U) =
\delta_{n,m} \; .}
This can be satisfied by choosing $\phi_n = \partial_U \psi_n/|m_n|$
for modes with
$m_n \neq 0$. For $m_n=0$, a formal extrapolation of the above analysis
leads to a
zero mode, $\phi_0 \sim \partial_u\psi_0\sim  1/H(U)$. In the
$D4-D8-\bar{D8}$ system,
an analogous mode was shown in \SakaiCN\  to be the massless
Nambu-Goldstone
boson associated with the breaking of $U(N_f)_L\times U(N_f)_R$ to
$U(N_f)_{\rm diag}$.
In our case,  in contrast to \SakaiCN, this putative zero mode is
neither orthogonal to the
other $\phi_n$, nor normalizable. The latter follows from the fact that
$H(U) \sim U$ at
large $U$ so that the integral
\eqn\divint{\int dU H(U) \phi_0(U)  \phi_0(U)\sim \int {dU\over H(U)}}
diverges logarithmically at large $U$. Thus, there is no normalizable
zero mode and hence no massless NG mode in the spectrum. This 
seems odd, since we have established above that the symmetry is 
broken and one expects this to lead to a NG boson on general grounds.
We will discuss the resolution of this puzzle below.

Another puzzling feature of the spectrum is the existence of a
normalizable zero mode for the
gauge field, corresponding to $\psi_0={\rm const}$ in \gaugemodes. It
satisfies \psieigen\ with
$m_n=0$ and from \psinorm\ is normalizable since $G(U) \sim 1/U^2$ at
large $U$. This seems
to suggest that the $U(N_f)$ symmetry on the $D6$-branes is gauged in
$1+1$ dimensions,
which is unexpected from the weak coupling  point of view. We will see
below that this mode
is not an eigenstate of the full quadratic action \expact.

To summarize, in the expansion \gaugemodes\ there is a normalizable
zero mode $\psi_0$ in the
expansion of $A_\mu$, but the expansion of $A_U$ involves a sum only
over non-zero modes
for which $\phi_n = \partial_U \psi_n/|m_n|$.  We thus see that
$A_U$ is pure gauge, $A_U = \partial_U \Lambda$ with $\Lambda = \sum_n
{1 \over |m_n|}
\pi^{(n)}(x) \psi_n(U)$. Making a gauge transformation by $\Lambda$ to
set $A_U=0$  and
substituting \gaugemodes\ into the action \dsixnext\  leaves us
with the two-dimensional action
\eqn\twodaction{\eqalign{
S_{D6}=- & \int d^2x\left[\frac{1}{4}B_{\mu\nu}^{(0)}B^{\mu\nu(0)} +
        \sum_{n\ge 1}\left(\frac{1}{4}B_{\mu\nu}^{(n)}B^{\mu\nu(n)} +
        \frac{1}{2}m_n^2B_{\mu}^{(n)}B^{\mu(n)}\right)\right] \cr
        - &  \int d^2 x \sum_{m,n}   \epsilon^{\mu \nu} {\cal
M}^{2}_{mn}B_\mu^{(n)} B_\nu^{(m)} }
}
where
\eqn\cscoef{{\cal M}^{2}_{nm} = - {\cal M}^{2}_{mn} = \frac{N_c}{2\pi}
\int dU \psi_n(U) \partial_U
\psi_m(U) \; .}
It is clear from \twodaction\ that the fields $B_\mu^{(n)}$ are not
mass eigenstates because of the off-diagonal mixing.  It is
difficult to compute the mass eigenstates explicitly since to do so
we would have to diagonalize the infinite-dimensional matrix with
diagonal components $m_n^2$ and off diagonal components ${\cal
M}^2_{n,m}$. However, there is no reason why the determinant of this
mass matrix should vanish, so this process presumably leaves us with
mass eigenstates all with non-zero masses. In particular, the
massless gauge field $B_\mu^{(0)}$ mixes with the massive modes via
\cscoef\ with $n=0$
\eqn\bzmix{{\cal M}^2_{0m} \propto {N_c \over 2 \pi} \int dU \partial_U
\psi_M(U) \; .}
It is easy to check that normalizable solutions of \psieigen\  approach
a non-zero constant
as $U \rightarrow \infty$, so the integral in \bzmix, and hence the
mixing of $B_\mu^{(0)}$
with the other modes, is non-zero.

One general observation that can be made about the spectrum is the
following. It was recently emphasized in \MyersQR\ that the
holographic distance-energy relation of \PeetWN\ coupled with the
consistency of holography for open and closed string modes implies a
universal scaling of meson masses  in the strongly coupled,
supergravity regime:
\eqn\mesonm{M_{\rm meson} \sim {m_f \over g_{\rm eff}(m_f)}}
where $m_f$ is the fermion mass and $g_{\rm eff}^2$ is the effective
't Hooft coupling evaluated at the scale $m_f$. In our model it is
given by
\eqn\geff{g_{\rm eff}^2 (m_f) \simeq g_5^2 N_c m_f\sim \lambda m_f~.}
Using the fact that $m_f\sim \lambda/L^2$, \mesonm\ leads to the
prediction that all meson masses are proportional to $M_{\rm meson}
\sim 1/L$ and are thus deeply bound since $M_{\rm meson} \ll m_f$.
To see that this is in fact true, note that the mass parameters
$m_n^2$ in \twodaction\ are determined by the eigenvalue equation
\psieigen. Writing it in terms of $\bar U = U/U_0$ gives
\eqn\resceigen{\bar U^2 \gamma^{-{1\over2}} \partial_{\bar U} \left(
\bar U \gamma^{-{1\over2}} \partial_{\bar U} \psi_n \right) = - {R^3
\over U_0} m_n^2 \psi_n \; .}
Since all the dependence on the parameters $\lambda$ and $L$ is now
contained in the factor $R^3/U_0 \sim L^2$, we see that the
eigenvalues $m_n^2$ are all proportional to $1/L^2$ and independent
of the 't Hooft coupling $\lambda$.

The mixing parameters ${\cal M}^{2}_{mn}$ \cscoef\ also have this
property.  The normalization
condition for $\psi_n$ \psinorm\ can be written as
\eqn\psnor{\int {d\bar U\over \bar U^2} \gamma^{1\over2} \bar \psi_n
\bar \psi_m = \delta_{n,m} }
where
\eqn\psidefn{\psi_n ={c \over L \sqrt{N_c}} \bar \psi_n}
for $c$ a constant, independent of $L$ and $\lambda$. Therefore,
${\cal M}^{2}_{m,n}$ \cscoef\  also scale like $1/L^2$.

To summarize, fluctuations of the gauge field lead to a spectrum of
massive vector fields in $1+1$ dimensions. The masses all scale like
$1/L$ and are determined by solving the eigenvalue equation
\psieigen\ and then diagonalizing the quadratic form in \twodaction.
A similar analysis leads to a spectrum of massive scalar and fermion
fields with masses that scale like $1/L$.

In contrast to the model of Sakai and Sugimoto \SakaiCN, we do not find
any massless
NG bosons in the spectrum, in spite of the apparent breaking of
$U(N_f)_L \times U(N_f)_R$
to the diagonal $U(N_f)$. The resolution of this puzzle was pointed out
in a closely related
context in \ItzhakiTU. We saw in section 3 (after eq. \linflow) that in
the weak coupling limit,
the NG bosons are displaced from the intersection of the branes by an
amount of order
$\lambda^{1/3}$ in the $U$ direction. Extrapolating this to strong
coupling we find that the
$U(N_f)$ degrees of freedom live in the transition region between the
near-horizon geometry
of the $D4$-branes and the asymptotically flat space far from the
branes. Therefore, they are
not visible in an analysis that restricts attention to the near-horizon
geometry. Note that this kind
of behavior is only possible in two dimensions where the NG bosons are
decoupled from the
massive physics. In higher dimensions similar behavior is familiar from
the analysis of the
center of mass modes in brane systems, which decouple in the
near-horizon limit.

The scalar, fermion and vector mesons obtained by studying small
fluctuations of the
D-branes are quite different from the bound states of the GN model
whose masses are
given by \masses. In particular, as is clear from our analysis, at weak
coupling the
color degrees of freedom living on the $D4$-branes are essentially
non-dynamical,
and their main effect is to produce the small attractive force between
the fermions
$q_L$, $q_R$. All the states \masses\ can then be thought of as bound
states of these
fermions.

On the other hand, at strong coupling, the degrees of freedom living
on the $D4$-branes (or, more properly, compactified $M5$-branes) are
strongly interacting, and the states with masses of order $1/L$ are
best thought of as containing the fermions $q_{L,R}$ as well as the
color degrees of freedom. They have the qualitative form
$q^\dagger\Phi^n q$, where $q$ stands for the fermions living at the
intersection, and $\Phi$ stands for all the bosonic and fermionic
degrees of freedom on the $D4$-branes. Therefore, these states do
not have a direct analog at weak coupling. It is interesting that
they are much lighter than the fermions themselves, like
Nambu-Goldstone bosons of a spontaneously broken symmetry.

To study the analogs of the GN states \masses\ at strong coupling,
one needs to consider long fundamental strings stretched between the
curved $D6$-branes and $U=0$. As mentioned above, a single string of
this sort is a massive fermion. A few such strings might form bound
states, like their weak coupling counterparts. This is an interesting
problem for future study.

\newsec{Finite temperature}

At zero temperature, the intersecting brane model of sections 2 -- 4
exhibits dynamical chiral symmetry breaking both at weak coupling
(discussed in section 3), where it reduces to the Gross-Neveu model,
and at strong coupling, where it is well described by probe
$D6$-brane dynamics in the near-horizon geometry of $D4$-branes
(section 4). In this section we will study this model at finite
temperature, and in particular analyze the finite temperature phase
transition at which chiral symmetry is restored. In the following
two subsections we do that in turn for weak and strong coupling. In
the last subsection we briefly address the interpolation between the
two regimes.

\subsec{The Gross-Neveu model at finite temperature}

The finite temperature properties of the Gross-Neveu model were
studied in \refs{\JacobsYS\HarringtonTF-\DashenXZ} and many
subsequent papers. Here we briefly review the results, which are
relevant for our intersecting D-brane system in the limit
$\lambda_{gn}\to 0$, $E\sim m_f$ discussed in section 3.  In this
limit the D-brane system is described by the local Lagrangian \lgn,
which can be written in a way similar to \tlreff:
\eqn\locsig{\SS_{\rm eff} = \int d^2x
\left(iq_L^\dagger\bar\sigma^\mu\partial_\mu q_L
+iq_R^\dagger\sigma^\mu\partial_\mu q_R+\sigma q_R^\dagger\cdot q_L
+\bar \sigma q_L^\dagger\cdot q_R-{N_c\over\lambda_{gn}}\sigma\bar
\sigma\right)~.} The expectation value of $\sigma$ is the mass of
the fermions $q_L,q_R$. To calculate it, it is convenient to
integrate out the fermions and minimize the resulting effective
potential for $\sigma$.

For zero temperature, the effective potential takes the form \GrossJV
\eqn\vvee{V_{\rm eff}={1\over\lambda_{gn}}\sigma\bar \sigma-\int
{d^2k\over (2\pi)^2}
\ln\left(1+{\sigma\bar\sigma\over k^2}\right)~.}
Stationary points of $V_{\rm eff}$ satisfy
\eqn\delvv{{\delta V_{\rm eff}\over \delta\bar
\sigma}={\sigma\over\lambda_{gn}}
-\int {d^2k\over (2\pi)^2}{\sigma\over k^2+\sigma\bar \sigma}=0~,}
which is equivalent to the gap equation \lmmfft\ discussed in section 3
in the local,
GN, limit. $\sigma=0$ is a local maximum of the potential \vvee; the
minimum
corresponds to the non-trivial solution of \delvv\ where chiral
symmetry is broken.

To study chiral symmetry breaking at finite temperature, we need to
evaluate the expectation value \eqn\fintt{\langle\sigma\rangle={{\rm
Tr} \sigma e^{-\beta H}\over {\rm Tr} e^{-\beta H}}} where $H$ is
the Hamiltonian and $1/\beta=T$ the temperature. This can be done by
studying the theory corresponding to \locsig\ on $\IR\times S^1$,
with the Euclidean time living on a circle of circumference $\beta$
with anti-periodic boundary conditions for the fermions $q_L$,
$q_R$. This amounts to replacing in \vvee, \delvv, \eqn\finitet{\int
{dk_0\over2\pi}\to {1\over\beta}\sum_{r\in Z+\half}} with $k_0=2\pi
r/\beta$ the momentum in the Euclidean time direction. Making this
replacement in \vvee\ and minimizing the effective potential with
respect to $\sigma$ gives the leading contribution to the free
energy $F(\beta)$. The value of $\sigma$ at the minimum gives the
expectation value \fintt.

The stationarity condition of $V_{\rm eff}$ is now
\eqn\coeffvv{{\delta V_{\rm eff}\over \delta\bar
\sigma}={\sigma\over\lambda_{gn}}\left[
1-{\lambda_{gn}\over\beta}\sum_{r\in Z+\half}\int_{-\Lambda}^\Lambda
{dp\over 2\pi}
{1\over\left({2\pi r\over\beta}\right)^2+  p^2+|\sigma|^2}\right]}
Following \DashenXZ\ one can replace the sum over $r$ by a contour
integral over energy,
\eqn\conint{{1\over\beta}\sum_{r\in Z+\half}\to \oint
{d\epsilon\over2\pi i}{1\over e^{\beta\epsilon}+1}~,}
where the contour surrounds the poles at $\beta\epsilon=2\pi r i$.
Deforming the contour
and picking up the residues of the poles of the r.h.s. of \coeffvv\ at
\eqn\reseps{\epsilon=\pm\epsilon_p=\pm\sqrt{p^2+|\sigma|^2}}
gives
\eqn\varfree{{\delta V_{\rm eff}\over \delta\bar
\sigma}={\sigma\over\lambda_{gn}}\left[
1-{\lambda_{gn}\over4\pi}\int_{-\Lambda}^\Lambda {dp\over \epsilon_p}
\tanh\left(\half\beta\epsilon_p\right)\right]~.}
The stationary points are $\sigma=0$ and the solution of
\eqn\nontriv{1={\lambda_{gn}\over2\pi}\int_0^\Lambda {dp\over
\epsilon_p}
\tanh\left(\half\beta\epsilon_p\right)~.}
As $\beta\to\infty$ we can approximate $\tanh\half\beta\epsilon_p\simeq
1$,
and \nontriv\ (with $\sigma(\beta\to\infty)=m_f$) reads
\eqn\perfint{1={\lambda_{gn}\over2\pi}\int_0^\Lambda {dp\over
\epsilon_p}\simeq
{\lambda_{gn}\over2\pi}\ln{2\Lambda\over m_f}~.}
Thus, the mass of the fermions at zero temperature is given by
\eqn\mftzero{m_f=2\Lambda e^{-{2\pi\over\lambda_{gn}}}~,}
which agrees with \mmferm\ after allowing for a different definition of
the cutoff $\Lambda$. 

For low temperature, \ie\ large but finite $\beta$, $\sigma=0$ is still
a local maximum.
Indeed, for $\sigma=0$, $\epsilon_p=|p|$ and the integral in \varfree\
is given by \DashenXZ:
\eqn\intres{\int_0^\Lambda {dp\over p}\tanh(\half\beta p)\simeq
\ln(1.14\beta\Lambda)~,}
where we assumed that $\beta\Lambda\gg 1$, as appropriate in the local
GN limit. Thus,
one has
\eqn\secder{{\delta^2V_{\rm
eff}(\sigma=0)\over\delta\sigma\delta\bar\sigma}=
{1\over\lambda_{gn}}\left(1-{\lambda_{gn}\over2\pi}\ln(1.14\beta\Lambda)\right)=
-{1\over2\pi}\ln(0.57\beta m_f)~,}
where in the last step we used \perfint. We see that $\sigma=0$ is a
maximum of $V_{\rm eff}$
for $\beta>\beta_c$ and a minimum for $\beta<\beta_c$. The critical
temperature is
\eqn\crittemp{T_c={1\over\beta_c}=0.57 m_f~.}
Below that temperature, the minimum of the effective potential is at a
non-zero value of $\sigma$, $\sigma=m_f(\beta)$, which corresponds to 
a non-trivial solution of \nontriv\ with chiral symmetry broken. As $T\to T_c$, 
$m_f(\beta)\to0$; for $T>T_c$ the only solution of \nontriv\ is $\sigma=0$, 
and chiral symmetry is restored. The phase transition at $T=T_c$ is second 
order since the order parameter $m_f(\beta)$ is continuous at the transition.

It should be noted that, as in section 3, the discussion above is
strictly speaking valid
only for infinite $N_c$. For finite $N_c$ the order parameter \fintt\
is not really constant
in space as was assumed in our analysis. Averaging over sufficiently
large distances,
this expectation value vanishes for all $T$. However the domains over
which it is roughly
constant grow with $N_c$. Thus, the large $N_c$ analysis is a useful
guide
to the behavior of the system for finite $N_c$ as well
(see \refs{\JacobsYS\HarringtonTF-\DashenXZ} for further discussion).

\subsec{Strong coupling analysis at finite temperature}

At strong coupling the dynamics of the intersecting branes is described
by studying probe
$D6$-branes in the near-horizon geometry of the $N_c$ $D4$-branes. To
study the system
at finite temperature we compactify Euclidean time on a circle of
circumference $\beta$, with
anti-periodic boundary conditions for the fermions. The resulting
metric and dilaton are given by
\eqn\dfourT{\eqalign{&ds^2=\left( U \over R\right)^{3\over2}
\left( f(U)(dx^0)^2+(dx^1)^2+(dx^4)^2\right)+
\left(U\over R\right)^{-{3\over2}}
\left(\frac{1}{f(U)} dU^2+U^2d\Omega_4^2\right)~, \cr
&e^\Phi=g_s\left(U\over R\right)^{3\over4}~,}}
with
\eqn\ffuu{f(U) = 1-{U_T^3\over U^3}~.}
This geometry is a continuation to Euclidean space of the
non-extremal $D4$-brane background. The non-extremality parameter
$U_T$ is related to the temperature via \eqn\timecond{
\beta=\frac{4\pi R^{3\over2}}{3U_T^{1\over2}} ~. }
For zero temperature one has $U_T=0$, and the solution \dfourT,
\ffuu\ reduces to \dfour.

To calculate the free energy of the intersecting brane theory we
need to solve for the shape of the $D6$-branes in the geometry
\dfourT. This can be done similarly to the closely related
$D4-D8-\bar{D8}$ case which was studied in
\refs{\AharonyDA,\ParnachevDN}.

The DBI action for the sixbranes is
\eqn\dbiT{S_{D6} = -T_6V_{1+1}V_4 R^{3\over2} \int dx^4 U^{5\over2}
\sqrt{f(U)+ \left(R\over U\right)^3 U'^2}~.}
The generalization of \conserv\ to non-zero temperature is
\eqn\eomT{
\frac{U^{5\over2}\sqrt{f(U)}}{\sqrt{1+\left(\frac{R}{U}\right)^3
\frac{U'^2}{f(U)}}} = U_0^{5\over2}\sqrt{f(U_0)} ~. }
$U_0$ is the lowest value of $U$ reached by the brane.

We need to solve \eomT\ for the shape of the branes $U(x^4)$, as in
section 4. There are again two possible solutions.  In one the $D6$
and $\bar{D6}$-branes wrap the smooth space labeled by $(U, x^0)$,
at two different values of $x^4$ a distance $L$ apart. In this
configuration the branes extend all the way to the Euclidean
horizon; thus $U_0=U_T$. The chiral $U(N_f)_L\times U(N_f)_R$
symmetry \gaugesym\ is preserved.

In the other solution, the $D6$ and $\bar{D6}$-branes are connected,
with $U=U(x^4)$,
and $U_0>U_T$. This solution breaks the chiral $U(N_f)_L\times
U(N_f)_R$ symmetry
to the diagonal $U(N_f)$.

To see which of the solutions has lower free energy we need to compare
the values of the
corresponding effective potentials
\eqn\endiffT{\eqalign{
\Delta {E}=E_{\rm straight}-E_{\rm curved} \sim &
  \int_{U_T}^{U_0} dU\left( U - 0 \right)
  + \int_{U_0}^\infty dU U\left[ 1 - \left(1-\frac{U_0^5 f(U_0)}{U^5
f(U)}\right)^{-{1\over2}} \right] \cr
\sim& \int_{y_T}^{1} dy \left( y - 0 \right)
  + \int_{1}^\infty dy y\left[ 1 - \left(1-\frac{f(1)}{y^5
f(y)}\right)^{-{1\over2}} \right]~, \cr
}}
where in the last line we introduced the dimensionless coordinate
$y=U/U_0$.

The numerical result for $\Delta E(T)$ is plotted in figure 2. For
low temperature, the curved solution is energetically preferred.
$\Delta E$ vanishes at a critical temperature corresponding to
$y_T^c=U_T^c/U_0 \approx 0.613$, above which the symmetric solution
becomes energetically preferred. Thus, the system exhibits a first
order phase transition at a temperature $T_c\simeq 1/L$. The
constant of proportionality can be fixed as follows. In the low
temperature phase, $L$, $U_T$ and $U_0$ are related by
\eqn\LT{ L =
2\frac{R^{3\over2}}{U_0^{1\over2}}\int_1^{\infty}dy\frac{y^{-{3\over2}}}{\sqrt{f(y)}\sqrt{\frac{f(y)}{f(1)}y^5-1}}\;.
}
At the critical temperature $y_T=y_T^c$ one finds $L \approx
1.097\frac{R^{3/2}}{U_0^{1/2}}$,
which implies
\eqn\ttcc{T_c\approx 0.205/L~.}
The phase transition at $T_c$ is first order since the fermion mass
jumps from a non-zero
value to zero as $T$ crosses $T_c$. The mass right below the transition
is equal to the energy of a string stretched
from the curved $D6$-branes to the Euclidean horizon at $U=U_T$,
\eqn\fermionM{
m_f = \frac{1}{2\pi}\int_{U_T}^{U_0}\sqrt{-g_{00}g_{UU}}dU =
\frac{1}{2\pi}(U_0-U_T)~.
}
Thus $\frac{m_f(T_c)}{m_f(0)} = 1 - y_T^c \approx 0.387$.

\ifig\etemp{Difference between the free energy of the straight and
curved solutions as a
function of temperature.}
{\epsfxsize3.0in\epsfbox{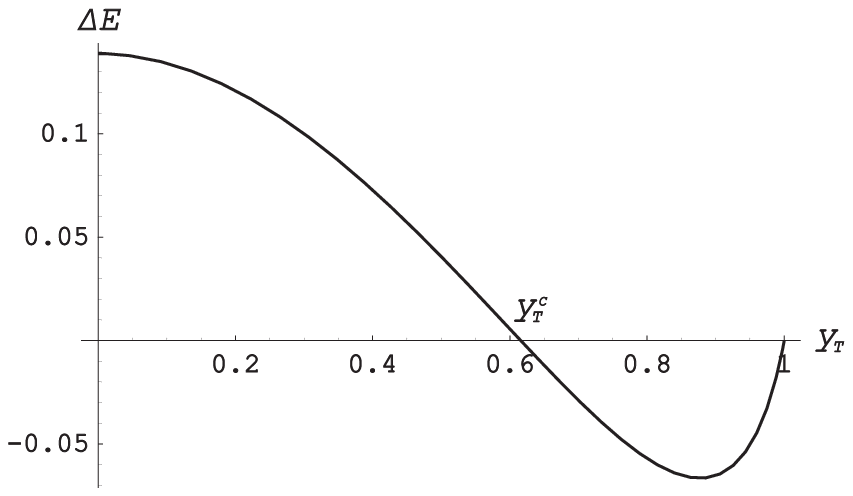}}

\subsec{Interpolating between weak and strong coupling}

In the last two subsections we discussed the finite temperature
phase transition in the limits $\lambda/L\to 0$ and
$\lambda/L\to\infty$.
It would be interesting to understand the interpolation between the
two
regimes.

The transition temperature, $T_c$, is given in the two limits by
\crittemp\
and \ttcc, respectively. A natural interpolation to all values of the
coupling
is
\eqn\tccoup{{T_c\over m_f}=H(m_fL)~.}
Here $m_fL$ is a measure of the coupling $\lambda/L$. It goes to zero
in the
GN limit and to infinity in the supergravity one (see \massll, \aaxx).
$H$ is a
function to be determined. Eq. \crittemp\ implies that
\eqn\hhoo{H(0)=0.57}
while \ttcc\ implies that
\eqn\hhii{H(x\to\infty)\simeq {0.205\over x}~.}
It is natural to expect that $H(x)$ varies smoothly and monotonically
with $x$.

Another interesting observable is the order parameter right below
the transition, \eqn\ordpar{{m_f(T_c)\over m_f(0)}=B(m_fL)~.} {}From
the GN analysis we know that \eqn\bboo{B(0)=0~,} \ie\ the symmetry
restoration transition is second order. We expect the transition to
remain second order for sufficiently small $\lambda/L$, assuming
that the physics at finite $\lambda/L$ is continuously connected to
that of the GN model.\foot{We thank O. Aharony for a discussion of
this issue.} The reason is that the effective potential \varfree\
behaves near the origin like \eqn\nearorig{V_{\rm eff}=
a_1|\sigma|^2+a_2|\sigma|^4+\cdots} where $a_1$ is given by \secder\
and changes sign at $T=T_c$ and $a_2$ is strictly positive. Small
$\lambda/L$ corrections to $a_1$ and $a_2$ cannot change the fact
that the transition happens at the point where $a_1$ changes sign
and is second order.

At strong coupling we found \eqn\bbii{B(\infty)=0.387~.} We expect
$B$ to vanish for $\lambda/L$ below some critical value, and then
smoothly interpolate between zero and \bbii\ as
$\lambda/L\to\infty$.

\newsec{Discussion}

In this paper we analyzed a model of intersecting $D4$ and
$D6$-branes and antibranes that has the interesting property that
the chiral symmetry of the original brane configuration \gaugesym\
is broken to the diagonal subgroup by the dynamics. The model
contains an adjustable parameter, $\lambda/L$, which governs the
strength of the coupling between left and right-moving fermions
living at different intersections. For weak coupling it reduces to
the Gross-Neveu model \GrossJV, a vector model which is exactly
solvable\foot{In fact this model is integrable at finite $N_c$ as
well.} at large $N_c$. For strong coupling it can be analyzed by
studying the DBI action for $D6$-branes in the near-horizon geometry
of $N_c$ $D4$-branes and again breaks chiral symmetry.

As one would expect, the strength of chiral symmetry breaking
increases with the coupling. For $\lambda/L\ll1$, the dynamically
generated fermion mass $m_f$ \mmferm\ is much smaller than the
natural scales in the problem, $1/L$ and $1/\lambda$. Indeed,
$m_fL$, $m_f\lambda$ go to zero as the coupling goes to zero.  For
$\lambda/L\gg1$ the fermion mass is of order  $\lambda/L^2$ and is
much larger than the other scales: $m_fL$, $m_f\lambda$ go to
infinity as the coupling goes to infinity (see \massll, \aaxx). The
strong interactions between the fermions and the degrees of freedom
living on the $D4$-branes give rise to strongly bound mesons whose
masses are much smaller than $m_f$ (their binding energy is thus
almost 100\%). These bound states can be studied by expanding the
DBI action of the sixbranes around the vacuum solution.

Another manifestation of the dependence of the strength of chiral
symmetry breaking on the coupling is the thermodynamic properties
described in section 5. For strong coupling, the transition is
strongly first order -- the fermion mass right below the transition
is of order $m_f$ (see \fermionM), which as mentioned above is large
in natural units. As the strength of the interaction decreases the
transition becomes weaker and it is second order at weak coupling
(as described in subsection 5.1).

There are many possible directions for further investigation of this
and related models.
The Gross-Neveu model has a spectrum of meson excitations with masses
given by \masses\
whose S-matrix is known due to integrability. In our setting, the GN
model is obtained in
the limit $\lambda/L\to 0$ described in section 3. It would be
interesting to see whether
the integrability persists for finite $\lambda/L$, and if so what one
can learn about the
model by using it. In particular it would be interesting to determine
the mass spectrum,
S-matrix and the thermodynamic functions $H$ \tccoup\ and $B$ \ordpar\
for all values of
the coupling.

Another issue involves the stability of the system  beyond the large
$N_c$ limit. In the full IIA string theory the brane configuration
of figure 1 is unstable. The $D6$ and $\bar{D6}$-branes attract each
other by massless closed string exchange and will start moving
towards each other if placed in this configuration. These
gravitational effects turn off in the limit $g_s\to 0$, and
therefore can be neglected in the 't Hooft limit that we took. It is
natural to ask what happens for finite $N_c$. We suspect that the
system is still stable although this deserves more careful study. On
the supergravity side, taking the near-horizon limit means that the
$D6$ and $\bar{D6}$-branes become infinitely separated near the
boundary at $U = \infty$. Therefore we expect the force per unit
worldvolume to go to zero, while the mass per unit worldvolume
remains constant. On the field theory side the dynamics is governed
by the $(2,0)$ theory compactified on a circle. The $D6$-branes are
interpreted as defects in this theory and we see no clear reason why
the introduction of defects should result in a drastic instability
of the theory.

There are many natural generalizations of our model. The brane
configuration of figure 1 contains a single stack of $D6$-branes all
at the same point in the $\IR^3$ labeled by (234), and $N_f$
antibranes at a distance $L$ in $\IR^3$. As a consequence, the
dynamics depends on a single parameter, $\lambda/L$. One could
generalize this  construction by putting the $D6$-branes and
antibranes at arbitrary points in $\IR^3$, and studying the dynamics
as a function of their relative positions. At weak coupling this
gives a generalization of the GN model, while at strong coupling it
gives a generalization of the DBI dynamics of section 4. In both
limits there is a rich phase structure with different patterns of
symmetry breaking depending on the positions of the branes and
antibranes in $\IR^3$. It would be interesting to analyze it.

Another generalization involves making the $SU(N_c)$ gauge fields
dynamical. This can be done by compactifying the directions (234) on
a torus of volume $V_{234}$. The resulting model is very similar to
the 't Hooft model of two-dimensional QCD \tHooftHX. The two-dimensional
 't Hooft coupling, which has units of energy squared and
sets the scale of the masses of mesons, is given by
\eqn\twodthooft{\lambda_2={\lambda\over V_{234}}~.} The dynamics
depends on the hierarchy of scales and the strength of the various
couplings. At weak coupling and large volume $V_{234}$, the scale of
confinement $\sqrt{\lambda_2}$ is much smaller than that of chiral
symmetry breaking $m_f$ \mmferm. Thus, at the scale of chiral
symmetry breaking one can neglect the dynamics of the two-dimensional 
gauge field, and the mass generation of the fermions
occurs as in section 3. At the much lower energy scale
$\sqrt{\lambda_2}$ the effect of confinement kicks in, and the
analysis of \tHooftHX\ is applicable (with fermion mass $m_f$). As
the volume $V_{234}$ decreases the physics changes, and for
$\lambda_2\gg m_f^2$ the gauge dynamics becomes strong at energy
scales at which the fermions are still massless.

Thus, in this model confinement and chiral symmetry breaking are
distinct phenomena and can occur at widely separated energies. This
property was noted in a four-dimensional intersecting brane system
in \AntonyanVW\ and is a general feature of such systems.

Other brane configurations with qualitatively similar dynamics are
obtained by changing the dimensionalities of the color and flavor
branes and that of the intersection. This leads to many interesting
models, including that of \refs{\SakaiCN,\AntonyanVW}, one that was
considered recently in \GaoUP, and many others. We will discuss some
features of such models in a companion paper \toap.

\bigskip\medskip\noindent
{\bf Acknowledgements:} We thank O. Aharony, N. Itzhaki, O. Lunin,
A. Parnachev, D. Sahakyan and K. Skenderis for discussions. DK
thanks the Weizmann Institute for hospitality during part of this
work. JH and DK thank the Aspen Center for Physics for providing a
supportive atmosphere during the completion of this work. The work
of EA and JH  was supported in part by NSF Grant No. PHY-0506630. DK
was supported in part by DOE grant DE-FG02-90ER40560. This work was
also supported in part by the National Science Foundation under
Grant 0529954. Any opinions, findings, and conclusions or
recommendations expressed in this material are those of the authors
and do not necessarily reflect the views of the National Science
Foundation.

\listrefs
\end